\documentclass[
	conference,
]{IEEEtran}

\IEEEoverridecommandlockouts 
\makeatletter
\def\markboth#1#2{\def\leftmark{\@IEEEcompsoconly{\sffamily}\MakeUppercase{\protect#1}}%
\def\rightmark{\@IEEEcompsoconly{\sffamily}\MakeUppercase{\protect#2}}}
\makeatother

\usepackage[amssymb]{SIunits}
\usepackage[cmex10]{amsmath} 
\usepackage{amssymb}
\usepackage[latin1]{inputenc}
\usepackage{xcolor}
\usepackage{cite}
\usepackage[pdftex]{graphicx}
\usepackage[caption=false, font=footnotesize]{subfig}
\usepackage{url}
\usepackage{xspace} 

\newenvironment{DIFnomarkup}{}{}

\graphicspath{%
	{../pstricks/}%
	{../mafig/}%
}%

\usepackage[
	shortcuts,
	nonumberlist,	
	acronym, 
	toc, 
	hyperfirst=true,
	section=section, 
	order=letter,
	numberedsection=nolabel 
]{glossaries}

\newglossary[slg]{los}{syi}{syg}{List of Symbols} 
\makeglossaries 
\glsdisablehyper 
\newglossarystyle{mylist}{
	\glossarystyle{list} 
}

\newglossarystyle{mysymbolstyle}{

	\renewcommand*{\glsgroupheading}[1]{}%
}

\newglossarystyle{myacronymstyle}{
  \glossarystyle{mysymbolstyle} %

}
\newcommand{\NewS}[5][\newcommand]{
	\newglossaryentry{symb:#2}{
		name=\ensuremath{#3},
		description={\nopostdesc #4}, 
		sort=sym#5,
		type=los,
	}
	\glsadd{symb:#2}
	\expandafter\def\csname #2\endcsname{\ensuremath{#3}} 
}

\newcommand{\NewC}[5][\glsadd]{
	\newglossaryentry{symb:#2}{
		name=\ensuremath{#3},
		description={\nopostdesc #4}, 
		sort=sym#5,
		type=los,
	}
	\expandafter\def\csname #2\endcsname{\ensuremath{#3}} 
}

\newcommand{\NewF}[6][\newcommand]{
	\newglossaryentry{symb:#2}{
		name=\ensuremath{#3\cdot#4},
		description={\nopostdesc #5}, 
		sort=sym#6,
		type=los,
	}
	\expandafter\def\csname #2\endcsname ##1{\ensuremath{#3{##1}#4}} 
}
\NewS{TXR}{R_0}{transmitted radius}{rx}
\NewS{txR}{r_0}{transmitted radius}{rx}

\NewS{TXP}{\Theta_0}{transmitted phase}{tx}
\NewS{txP}{\theta_0}{transmitted phase}{tx}

\NewS{RXR}{R}{received radius}{ry}
\NewS{rxR}{r}{received radius}{ry}

\NewS{RXP}{\Theta}{received phase}{ty}
\NewS{rxP}{\theta}{received phase}{ty}

\NewS{TXRdet}{\hat{R}_0}{transmitted radius}{rx}

\NewS{estX}{\hat{X}}{estimated symbol}{xh}
\NewS{esttxR}{\hat{R}_x}{estimated transmitted radius}{rxe}
\NewS{esttxP}{\hat{\Theta}_x}{estimated transmitted phase}{txe}

\NewS{NLPN}{\Phi_{\text{NL}}}{nonlinear phase noise}{phi}
\NewS{Len}{{L}}{total fiber length}{L}
\NewS{Power}{{P}}{input power}{L}

\NewS{X}{X}{transmitted symbol}{x}
\NewS{Y}{Y}{received symbol}{y}
\NewS{setX}{\mathcal{X}}{signal constellation}{xs}
\NewS{setAPSK}{\setX_{\text{APSK}}}{APSK signal set}{xsapsk}

\NewS{rad}{r}{radius}{r}
\NewS{ppr}{l}{points per ring}{l}
\NewS{phaseoffset}{\phi}{phase offset per ring}{p}
\NewS{NumRings}{N}{number of rings in the APSK constelllation}{N}
\NewS{RD}{\bold{r}}{radii distribution}{rr}

\NewS{NumSpans}{K}{number of spans}{k}

\NewS{SEP}{\text{SEP}}{symbol error probability}{ty}

\NewS{labeling}{\mathbb{L}}{labeling}{l}
\NewS{BRGC}{\mathbb{G}}{binary reflected gray code}{gc}
\NewS{ODP}{\otimes}{ordered direct product}{odp}

\NewS{Reg}{\mathcal{R}}{compact bounding region}{reg}

\NewS{imag}{\jmath}{imaginary unit}{j} 

\NewC{Code}{\mathcal{C}}{code}{C:code}
\NewC{vecI}{\boldsymbol{I}}{identity matrix}{i}
\NewC{vecZero}{\boldsymbol{0}}{all-zero vector}{0z}
\NewC{xor}{\oplus}{exclusive or}{0excl}
\NewC{SNR}{\text{SNR}}{signal to noise ratio}{s}
\NewC{mod}{\;\operatorname{mod}\,}{modulo operation}{m}	
\NewC{cconv}{\circledast}{circular convolution}{0}	

\NewC{natural}{\mathbb{N}}{set of natural numbers}{n} 
\NewC{real}{\mathbb{R}}{set of real numbers}{rs}
\NewC{rational}{\mathbb{Q}}{set of rational numbers}{q}
\NewC{integer}{\mathbb{Z}}{set of integer numbers}{zzzz}
\NewC{complex}{\mathbb{C}}{set of complex numbers}{c}

\NewC{Rn}{\mathbb{R}^n}{real Euclidean $n$-dimensional space}{rsn}
\NewC{GF}{\mathbb{F}_2}{Galois field of size two}{f}

\NewC{N}{\mathcal{N}}{normal distribution}{no}
\NewC{Norm}{\mathcal{N}}{normal distribution}{N}
\NewC{Unif}{\operatorname{Unif}}{uniform distribution}{uz}
\NewF{norm}{||}{||}{Euclidean norm}{0e}
\NewF{Qla}{Q_{\Lambda}(}{)}{nearest neighbor lattice quantizer}{quant}
\NewF{Qco}{Q_{\Code}(}{)}{binary quantizer with respect to a linear code $\Code$}{quant2}
\NewF{E}{\mathbb{E}\left[}{\right]}{expectancy operator}{e}
\NewF{bef}{\operatorname{H}(}{)}{binary entropy function}{he}
\NewF{Pr}{\operatorname{Pr}\left[}{\right]}{probability of an event}{pr}

\newcommand{\IE}{i.e., } 
\newcommand{\EG}{e.g., } 

\newcommand{\define}{\triangleq}
\newcommand{\vect}[1]{\boldsymbol{#1}}

\renewcommand{\epsilon}{\varepsilon}
\renewcommand{\phi}{\varphi}

\makeatletter
\newcommand{\vast}{\bBigg@{5}}
\makeatother

\definecolor{shadecolor}{rgb}{0.97,0.97,0.97}%
\definecolor{framecolor}{rgb}{0,0,0}%

\newcommand{\abbr}[1]{{#1}}				

\makeatletter
\let\aclOLD=\acl
\renewcommand{\acl}[1]{%
  \begingroup    
  \let\@@underline=\relax
  \aclOLD{#1}%
  \endgroup
}
\makeatother

\newcommand{\NewA}[3]{
	\newacronym{#1}{#2}{#3}
}

\NewA{af}{AF}{\abbr{a}mplify-and-\abbr{f}orward}
\NewA{apsk}{APSK}{\abbr{a}mplitude \abbr{p}hase-\abbr{s}hift \abbr{k}eying}
\NewA{ask}{ASK}{\abbr{a}mplitude-\abbr{s}hift \abbr{k}eying}
\NewA{ase}{ASE}{\abbr{a}mplified \abbr{s}pontaneous \abbr{e}mission}
\NewA{awgn}{AWGN}{\abbr{a}dditive \abbr{w}hite \abbr{G}aussian \abbr{n}oise}
\NewA{bep}{BEP}{\abbr{b}it \abbr{e}rror \abbr{p}robability}
\NewA{ber}{BER}{\abbr{b}it \abbr{e}rror \abbr{r}ate}
\NewA{qap}{QAP}{\abbr{q}uadratic \abbr{a}assignment \abbr{p}roblem}
\NewA{bicm}{BICM}{\abbr{b}it-\abbr{i}nterleaved \abbr{c}oded \abbr{m}odulation}				
\NewA{bpsk}{BPSK}{\abbr{b}inary \abbr{p}hase-\abbr{s}hift \abbr{k}eying}				
\NewA{bsc}{BSC}{\abbr{b}inary \abbr{s}ymmetric \abbr{c}hannel}				
\NewA{brgc}{BRGC}{\abbr{b}inary \abbr{r}eflected \abbr{G}ray \abbr{c}ode}				
\NewA{cf}{CF}{\abbr{c}haracteristic \abbr{f}unction}
\NewA{csit}{CSIT}{\abbr{c}annnel \abbr{s}tate \abbr{i}nformation at the \abbr{transmitter}}
\NewA{csi}{CSI}{\abbr{c}annnel \abbr{s}tate \abbr{i}nformation}
\NewA{df}{DF}{\abbr{d}ecode-and-\abbr{f}orward}
\NewA{fd}{FD}{\abbr{f}ull-\abbr{d}uplex}
\NewA{fft}{FFT}{\abbr{f}ast \abbr{F}ourier \abbr{t}ransform }
\NewA{hd}{HD}{\abbr{h}alf-\abbr{d}uplex}
\NewA{iid}{IID}{\abbr{i}ndepend and \abbr{i}dentically \abbr{d}istributed }
\NewA{isi}{ISI}{\abbr{i}nter\abbr{s}ymbol \abbr{i}nterference }
\NewA{lb}{LB}{\abbr{l}ower \abbr{b}ound}
\NewA{map}{MAP}{\abbr{m}aximum \abbr{a} \abbr{p}osteriori}
\NewA{mf}{MF}{\abbr{m}odulo-and-\abbr{f}orward}
\NewA{mlan}{MLAN}{\abbr{m}odulo-\abbr{l}attice \abbr{a}dditive \abbr{n}oise}
\NewA{ml}{ML}{\abbr{m}aximum \abbr{l}ikelihood}
\NewA{mmse}{MMSE}{\abbr{m}inimum \abbr{m}ean \abbr{s}quare \abbr{e}rror}
\NewA{nlpc}{NLPC}{\abbr{n}onlinear \abbr{p}hase \abbr{c}ompensation}
\NewA{nlpn}{NLPN}{\abbr{n}onlinear \abbr{p}hase \abbr{n}oise}
\NewA{nc}{NC}{\abbr{n}etwork \abbr{c}oding}
\NewA{ofmd}{OFDM}{\abbr{o}rthogonal \abbr{f}requency-\abbr{d}ivision \abbr{m}ultiplexing }			
\NewA{dp}{DP}{\abbr{d}ual-\abbr{p}olarization}
\NewA{pam}{PAM}{\abbr{p}ulse \abbr{a}mplitude \abbr{m}odulation}
\NewA{pdf}{PDF}{\abbr{p}robability \abbr{d}ensity \abbr{f}unction}
\NewA{plnc}{PNC}{\abbr{p}hysical-layer \abbr{n}etwork \abbr{c}oding}
\NewA{psk}{PSK}{\abbr{p}hase-\abbr{s}hift \abbr{k}eying}
\NewA{pmd}{PMD}{\abbr{p}olarization \abbr{m}ode \abbr{d}ispersion}
\NewA{pdm}{PDM}{\abbr{p}olarization-\abbr{d}ivision \abbr{m}ultiplexing}
\NewA{qam}{QAM}{\abbr{q}uadrature \abbr{a}mplitude \abbr{m}odulation}
\NewA{sqp}{SQP}{\abbr{s}equential \abbr{q}uadratic \abbr{p}rogramming}
\NewA{rd}{RD}{\abbr{r}adii \abbr{d}istribution}
\NewA{sep}{SEP}{\abbr{s}ymbol \abbr{e}rror \abbr{p}robability}
\NewA{ser}{SER}{\abbr{s}ymbol \abbr{e}rror \abbr{r}ate}
\NewA{si}{SI}{\abbr{s}ide \abbr{i}nformation}
\NewA{sp}{SP}{\abbr{s}ingle-\abbr{p}olarization}
\NewA{sanr}{SNR}{\abbr{s}ignal-to-(additive-)\abbr{n}oise \abbr{r}atio}
\NewA{snr}{SNR}{\abbr{s}ignal-to-\abbr{n}oise \abbr{r}atio}
\NewA{snlse}{sNLSE}{\abbr{s}tochastic \abbr{n}onlinear \abbr{S}chr\"odinger \abbr{e}quation}
\NewA{stwrc}{sTRC}{\abbr{s}eparated \abbr{t}wo-way \abbr{r}elay \abbr{c}hannel}
\NewA{stwtrc}{sTTRC}{\abbr{s}eparated \abbr{t}wo-way \abbr{t}wo-\abbr{r}elay \abbr{c}hannel}
\NewA{ts}{TS}{\abbr{t}wo-\abbr{s}tage}
\NewA{twrc}{TRC}{\abbr{t}wo-way \abbr{r}elay \abbr{c}hannel}
\NewA{twtrc}{TTRC}{\abbr{t}wo-way \abbr{t}wo-\abbr{r}elay \abbr{c}hannel}
\NewA{wdm}{WDM}{\abbr{w}avelength \abbr{d}ivision \abbr{m}ultiplexing}

\newacronym[%
	longplural={binary erasure channels},%
	shortplural={BECs}%
]{bec}{BEC}{binary erasure channel}%

\NewA{scldpc}{SC-LDPC}{spatially coupled low-density parity check}

\NewA{vn}{VN}{variable node}
\NewA{cn}{CN}{check node}

\NewA{de}{DE}{density evolution}

\NewA{sc}{SC}{spatially-coupled}

\NewA{ldpc}{LDPC}{low-density parity check}

\NewA{bp}{BP}{belief propagation}

\NewA{dm}{DM}{dispersion-managed}

\NewA{spm}{SPM}{self-phase modulation}
\NewA{xpm}{XPM}{cross-phase modulation}
\NewA{fwm}{FWM}{four-wave-mixing}

\NewA{ixpm}{IXPM}{intrachannel cross-phase modulation}
\NewA{ifwm}{IFWM}{intrachannel four-wave-mixing}
\NewA{ssfm}{SSFM}{split-step Fourier method}

\newcommand{\rDegree}{\ensuremath{d_{\text{c}}}}
\newcommand{\lDegree}{\ensuremath{d_{\text{v}}}}
\newcommand{\coupling}{\ensuremath{{w}}}

\newcommand{\VNs}{\glspl{vn}\xspace}
\newcommand{\CNs}{\glspl{cn}\xspace}
\newcommand{\BECs}{\glspl{bec}\xspace}

\newcommand{\precision}{\delta}
\newcommand{\pe}{p_{\text{tar}}}
\newcommand{\lmax}{l_\text{max}}
\newcommand{\lsuccess}{l_\text{s}(\mathbf{A}, \bar{\eps})}

\newcommand{\eps}{\varepsilon}

\newcommand{\Aunif}{\mathbf{A}_\text{uni}}

\newcommand{\Aopt}{\ensuremath{\mathbf{A\!{ }^*}}}
\begin{document}

\begin{DIFnomarkup}

\title{Optimized Bit Mappings for Spatially Coupled LDPC Codes over
Parallel Binary Erasure Channels}


\author{
	\IEEEauthorblockN{
	Christian Häger\IEEEauthorrefmark{2},
	Alexandre Graell i Amat\IEEEauthorrefmark{2},
	Alex Alvarado\IEEEauthorrefmark{3}, 
	Fredrik Brännström\IEEEauthorrefmark{2}, and
	Erik Agrell\IEEEauthorrefmark{2}
	\thanks{This work was partially funded by
	the Swedish Research Council under grant \#2011-5961 
	and by the European Community's Seventh's Framework Programme (FP7/2007-2013) under grant agreement No.~271986. The calculations were performed in
	part on resources provided by the Swedish National Infrastructure
	for Computing (SNIC) at C3SE.} }
	\IEEEauthorblockA{\IEEEauthorrefmark{2}%
	Department of Signals and Systems,
	Chalmers University of Technology,
	Gothenburg, Sweden}
	\IEEEauthorblockA{\IEEEauthorrefmark{3}%
	Department of Engineering, University of Cambridge, UK
	}
	\{christian.haeger, alexandre.graell, fredrik.brannstrom, agrell\}@chalmers.se, alex.alvarado@ieee.org
}

\maketitle

\end{DIFnomarkup}

\begin{abstract}




	In many practical communication systems, one binary encoder/decoder
	pair is used to communicate over a set of parallel channels.
	Examples of this setup include multi-carrier transmission,
	rate-compatible puncturing of turbo-like codes, and \gls{bicm}. A
	bit mapper is commonly employed to determine how the coded bits are
	allocated to the channels. In this paper, we study spatially coupled
	low-density parity check codes over parallel channels and optimize
	the bit mapper using \gls{bicm} as the driving example. For
	simplicity, the parallel bit channels that arise in \gls{bicm} are
	replaced by independent \glspl{bec}. For two parallel \glspl{bec}
	modeled according to a 4-PAM constellation labeled by the binary
	reflected Gray code, the optimization results show that the decoding
	threshold can be improved over a uniform random bit mapper, or,
	alternatively, the spatial chain length of the code can be reduced
	for a given gap to capacity. It is also shown that for rate-loss
	free, circular (tail-biting) ensembles, a decoding wave effect can
	be initiated using only an optimized bit mapper.
	

\end{abstract}


\glsresetall

%
%
%
%
%
%

\section{Introduction}
\label{sec:introduction}

Spatial coupling of regular \gls{ldpc} codes has emerged as a powerful
technique to construct capacity-achieving codes for many communication
channels using iterative \gls{bp} decoding \cite{Lentmaier2005,
Kudekar2011}.  
In this paper, we apply spatially coupled LDPC
(SC-LDPC)\glsunset{scldpc} codes to a system where communication takes
place over a set of parallel channels. Parallel channels are
frequently encountered in practical scenarios, including multi-carrier
transmission and rate-compatible puncturing of turbo-like codes
\cite{Sason2007}. Our main motivation comes from the application of
\gls{scldpc} codes to \gls{bicm} systems, which are often analyzed
under the assumption of equivalent parallel binary-input channels (or
simply bit channels) \cite[Sec.~2-C]{Caire1998}. 


For a fixed code and channel characteristics, an important problem is
how to allocate the coded bits to the channels. This allocation is
performed by a so-called bit mapper\footnote{In the
literature, the term ``bit interleaver'' is also frequently used.}.
Our focus is on the asymptotic behavior and we are interested in
optimizing the bit mapper in terms of the decoding threshold. The
decoding threshold divides the parameter range used to characterize
the quality of the set of parallel channels into a region where
reliable decoding is possible and where it is not. Under the
assumption of infinite codeword length, \gls{de} can be applied in
order to find the threshold for \gls{ldpc} codes and \gls{bp} decoding
\cite{Richardson2001}. 


\Gls{scldpc} codes have been studied in the context of \gls{bicm}
systems in \cite{Schmalen2012} and \cite{Yedlaa} assuming a uniform
random bit mapping. 
Many authors have studied the optimization of bit mappers for
irregular \gls{ldpc} code ensembles. For example, in \cite{Cheng2012,
Nowak2012}, the authors use extrinsic information transfer charts to
find approximate decoding thresholds and subsequently optimize bit
mappers, where in \cite{Nowak2012} special attention is payed to the
short block length regime. In \cite{Richter2007}, a downhill algorithm
is used to find optimized bit mappings for two different
\gls{ldpc} codes. In \cite{Gong2011}, the authors substitute the
parallel \gls{bicm} bit channels by binary-input \gls{awgn} channels
and optimized bit mappers are found for the codes and modulations
specified in the DVB-T2 standard. Bit mappers with a simple
implementation structure are designed in \cite{Lei2009}. Some authors
have also devised heuristic bit mapping strategies \cite{Ryan2005}.
Furthermore, there exists a substantial amount of literature dealing
with both code optimization for a fixed bit mapping (\EG
\cite{Hou2003}) as well as the joint optimization of the code ensemble
and the bit mapper (\EG \cite{Durisi2006}).

In this paper, we take a similar approach as in \cite{Lei2009} and
substitute the parallel bit channels that arise in \gls{bicm} with
independent \BECs. This is justified by the observation that the
decoding threshold is mainly determined by the mutual information of
the channel rather than the channel details, see the discussion in
\cite[Sec. I]{Lei2009}. Compared to \cite{Gong2011}, where the
parallel channels are approximated as binary-input AWGN channels, the
numerical complexity of the \gls{de} equations is greatly simplified
when studying \BECs. 


The results in this paper are for a scenario with two parallel \BECs
modeled according to a 4-PAM constellation labeled by the \gls{brgc}
and we also briefly discuss the generalization to an arbitrary number
of channels. Optimized bit mappers are found for \gls{scldpc} code
ensembles with a two-sided termination boundary
\cite[Sec.~2-B]{Kudekar2011} as well as circular (tail-biting)
ensembles \cite[Sec.~V]{Kudekar2010a}. For the two-sided ensembles, it
is shown that the decoding threshold can be improved, or equivalently,
the spatial chain length can be reduced for a given gap to capacity
compared to a uniform random bit mapper.  Circular ensembles on the
other hand have a decoding behavior resembling that of regular
uncoupled ensembles due to the absence of a termination boundary. We
show that by using an optimized bit mapper, the different qualities of
the parallel channels can be exploited to obtain a decoding wave
effect as for two-sided ensembles, \IE the channels are effectively
used to induce a termination boundary.

\section{System Model}

\begin{figure}[t]
	\begin{center}
		\includegraphics{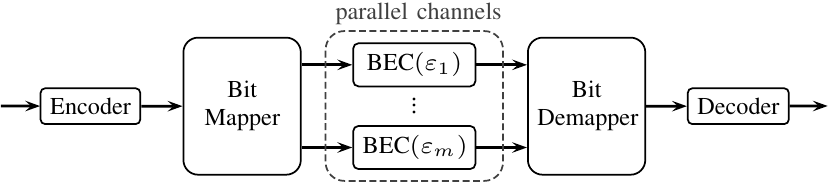}
	\end{center}
	\caption{Block diagram of the considered system model.}
	\label{fig:block_diagram}
\end{figure}

We consider a communication system where one binary encoder/decoder
pair is used to communicate over a set of parallel channels and a bit
mapper determines the allocation of coded bits to the channels. A
block diagram of the system model is shown in
Fig.$\!$ \ref{fig:block_diagram}. In the following, the individual blocks
are described in more detail.

\subsection{Parallel Channels}

Many practical transmission scenarios can be modeled as a set of
parallel channels and we take \gls{bicm} as an example throughout this
paper. To this end, consider the real \gls{awgn} channel $Y = X + N$,
where $X \in \mathcal{X}$ is the channel input taking on values from a
discrete signal constellation $\mathcal{X}$ and $N \sim \mathcal{N}(0,
1)$. If we label each element in the constellation with a unique
binary string of length $m = \log_2 |\mathcal{X}|$, then,
conceptually, we may view this setup as having $m$ parallel bit
channels from $B_i$ to $Y$, where $B_i$, $1\leq i \leq m$, denotes the
$i$th bit in the binary strings (counting from left to right)
\cite[Sec.~2-C]{Caire1998}. Each of these bit channels can be
characterized by an individual channel quality parameter $\alpha_i
\define 1 - I(B_i; Y)$ ranging from 0 (perfect channel) to 1 (useless
channel). The mutual information $I(B_i; Y)$ is commonly parameterized
by the \gls{snr} and depends on the signal constellation as well as
the binary labeling \cite{Caire1998}. 

\begin{figure}[t]
	\centering
	\quad\subfloat[4-PAM]{\includegraphics{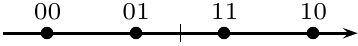}}
	\qquad
	\subfloat[8-PAM]{\includegraphics{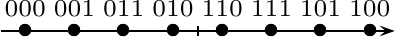}}
	\caption{Two PAM constellations labeled by the \gls{brgc}. }
	\label{fig:constellations}
\end{figure}

\begin{figure}[t]
	\centering

	\subfloat[4-PAM, BRGC]{\includegraphics{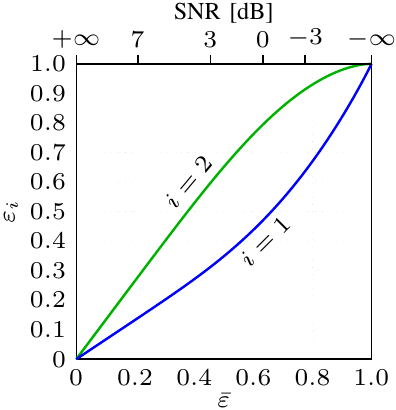}}
	\qquad
	\subfloat[8-PAM, BRGC]{\includegraphics{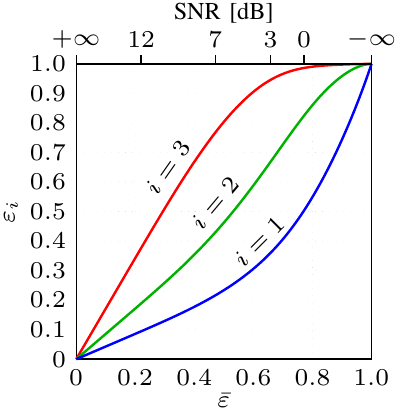}}

	\caption{Individual erasure probabilities $\eps_i$ plotted over
	$\bar{\eps}$. }
	\label{fig:channel_model}
\end{figure}

For simplicity, we replace these bit channels by parallel, independent
\glspl{bec} with erasure probabilities $\eps_i = \alpha_i$. In
Fig.~\ref{fig:constellations}, two PAM constellations labeled by the
\gls{brgc} are shown, and in Fig.\ \ref{fig:channel_model} we plot the
corresponding $\eps_i$ as a function of the average erasure
probability $\bar{\eps} = (\sum_{i=1}^{m} \eps_i)/m$. Note that
$\bar{\eps}$ (ranging from 0 to 1) is implicitly parameterized by the
\gls{snr} (ranging from $+\infty$ dB to $-\infty$ dB), indicated by
the top scale in Fig.~\ref{fig:channel_model}, and this
parameterization is different for the two constellations. Henceforth,
$\bar{\eps}$ is used as the parameter to characterize the overall
quality of the set of parallel channels (cf.\ Fig.\
\ref{fig:block_diagram}) and the correspondence between $\bar{\eps}$
and the individual channel qualities is according to Fig.\
\ref{fig:channel_model}. 



\subsection{Encoder and Decoder}

We focus on the two-sided and circular spatially coupled $(\lDegree,
\rDegree, L, \coupling)$ code ensembles, where $\lDegree$ and
$\rDegree$ denote the \gls{vn} and \gls{cn} degrees, $L$ the spatial
chain length, and $\coupling$ is a ``smoothing'' parameter.  
\begin{figure}[t]
	\begin{center}
		\includegraphics[width=8.5cm]{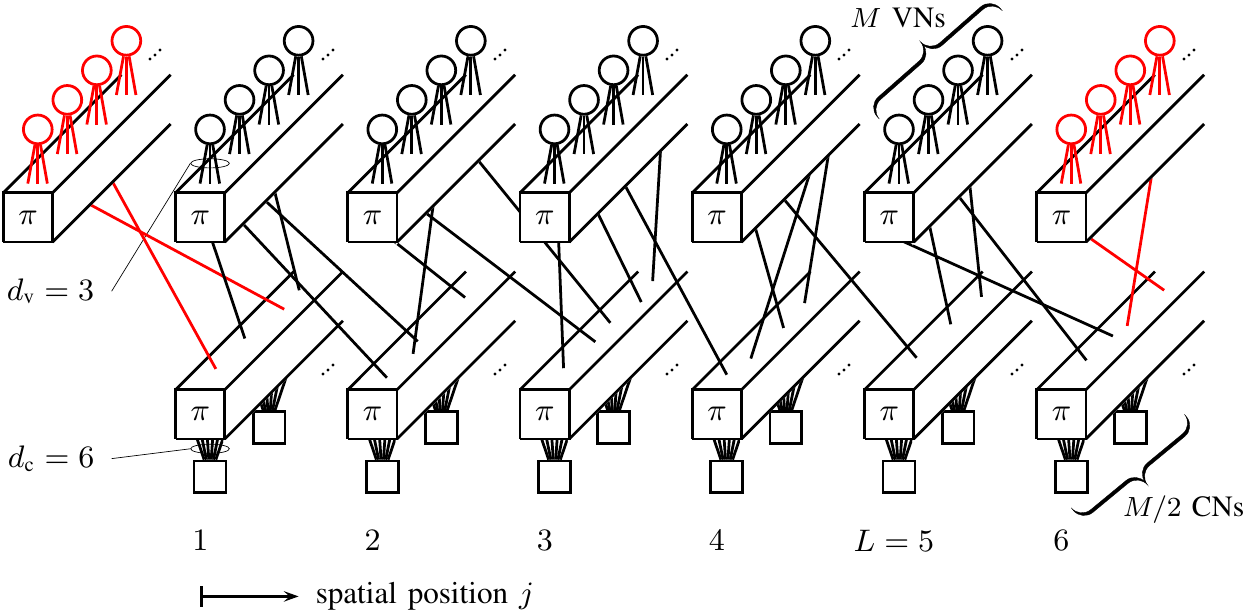}
	\end{center}
	\caption{Graphical representation of the Tanner graph for the
	two-sided $(3,6,5,2)$ spatially coupled ensemble. Known \VNs are shown
	in red.} 
	\label{fig:sc_ensemble_simple}
\end{figure}
The construction of the two-sided ensemble is explained in detail in
\cite[Sec.  2-B]{Kudekar2011} and extended to circular ensembles in
\cite[Sec.  V]{Kudekar2010a}. For completeness, we review the
construction with the help of the example depicted in Fig.\
\ref{fig:sc_ensemble_simple}, where $\lDegree=3$, $\rDegree=6$, $L=5$,
and $\coupling=2$, starting with the two-sided case. $M$ \VNs are
placed at each spatial position $1$ to $L$ and $\lDegree M/\rDegree$
\CNs are placed at each position $1$ to $L+\coupling-1$.  For the
asymptotic case, \IE infinite codeword length, it is assumed that $M
\to \infty$. The red circles in Fig.\ \ref{fig:sc_ensemble_simple}
correspond to \emph{known} \VNs which are initialized to zero erasure
probability and placed at positions $-w+2$ to $0$ and $L+1$ to
$L+w-1$. The connections between \VNs and \CNs are as follows. It is
assumed that the $\lDegree M$ edges originating from \VNs at position
$j$ are uniformly and independently distributed to \CNs at positions
$j$ to $j+\coupling-1$, whereas the $\lDegree M$ edges from \CNs at
position $j$ are assumed to be uniformly and independently distributed
to \VNs at positions $j-\coupling+1$ to $j$.  In the figure, this is
represented by the interleaver blocks that uniformly spread out the
edges from the \VNs and \CNs. 
For circular ensembles, one can apply the same construction as above,
but all position indices are now interpreted modulo $L$ and no known
\VNs are present \cite[Sec. V]{Kudekar2010a}. For the example in Fig.\
\ref{fig:sc_ensemble_simple}, this would correspond to removing the
red nodes and edges and appropriately connecting the \VNs at position
5 to the \CNs at position 1. Also, no \CNs are placed at
position 6 due to the modulo indexing.

The design rate $R$ of the two-sided ensemble is given by \cite[Lemma
3]{Kudekar2011} 
\begin{align}
	R = 1 - \frac{\lDegree}{\rDegree} - \frac{\lDegree}{\rDegree}
	\frac{\coupling + 1 - 2 \sum_{i=0}^{\coupling} \left(
	\frac{i}{\coupling} \right)^{\rDegree}}{L},
	\label{eq:designrate}
\end{align}
and it can be seen that there is a rate loss with respect to the
design rate $1-\lDegree/\rDegree$ of the underlying regular ensemble.
This is due to the termination boundary and the fact that a certain
fraction of \CNs is only connected to known \VNs. For the circular
ensemble, the rate loss is zero since all \VNs and \CNs are fully
connected. 



From the above definition of the ensemble, one can develop \gls{de}
equations that describe the temporal and spatial evolution of the
\gls{vn} erasure probabilities when performing \gls{bp} decoding under
the assumption that $M \to \infty$. In this case, the \gls{vn} erasure
probabilities are given by \cite[eq. (7)]{Kudekar2011}
\begin{align}
	p^{(l+1)}_j = \eps^{j} \left(\frac{1}{\coupling}\sum_{a=0}^{\coupling} \left( 1 - \left(1 -
	\frac{1}{\coupling}\sum_{b=0}^{\coupling} p^{(l)}_{j+a-b} \right)^{\rDegree-1} \right) \right)^{\lDegree-1}
	\label{eq:de}
\end{align}
for $-\coupling+2 \leq j \leq L +\coupling-1$, where $l$ denotes the
iteration number and $\eps^j$ the input erasure probability for the
\VNs at position $j$. The initial conditions for the two-sided
ensemble are $p_j^{(0)} = \eps^j$, where $\eps^j=0$ for $j<1$ and
$j>L$ due to the known \VNs. For circular ensembles, $j \in \{1, \cdots,
L\}$ and the index arithmetic in \eqref{eq:de} is performed modulo $L$
\cite[Sec.  V]{Kudekar2010a}. 

\subsection{Bit Mapper and Demapper}

For the considered scenario, the \gls{de} equations \eqref{eq:de} are,
in principle, the same as for the well-studied case with only one
\gls{bec}. The only difference is that the input erasure probabilities
can be different for each spatial position, \IE one may think of the
\VNs at different positions belonging to different equivalence
classes. One possible way to describe the assignment of channels to
\gls{vn} classes is via a matrix $\mathbf{A} = [a_{i,j}] \in
\mathbb{R}^{m \times L}$,
where $a_{i,j}$, $0 \leq a_{i,j} \leq 1$ $\forall i, j$, denotes the
fraction of \VNs from position $j$ to be sent over the $i$th
\gls{bec}. If we collect the individual channel erasure probabilities
in a vector $\vect{\eps} \define (\eps_1, \dots, \eps_m)$, then,
multiplying $\vect{\eps}$ by $\mathbf{A}$ leads to a vector $(\eps^1,
\eps^2, \dots, \eps^L)$ with the input erasure probabilities of the
$L$ \gls{vn} classes. Each input erasure probability is thus a
weighted average of the channel erasure probabilities. In order to
have a valid assignment, all columns in $\mathbf{A}$ have to sum up to
one and all rows in $\mathbf{A}$ have to sum up to $L/m$. The first
condition ensures that all \VNs are assigned to a channel, while the
second condition ensures that all parallel channels are used equally
often\footnote{The constraints on $\mathbf{A}$ may be
different for scenarios other than \gls{bicm}.}. The set of valid
assignment matrices that fulfill the above conditions is denoted by
$\mathcal{A}^{m \times L} \subset \mathbb{R}^{m \times L}$. 



\section{Decoding Threshold and Potential Gains}

For a fixed bit mapper, \IE for a fixed assignment matrix
$\mathbf{A}$, the decoding threshold $\bar{\eps}^*(\mathbf{A})$ is
defined as the largest $\bar{\eps} \in [0, 1]$ such that $\lim_{l \to
\infty} {p}^{(l)}_j = {0}$, $1 \leq j \leq L$, cf. \eqref{eq:de}. This
condition corresponds to successful decoding, \IE all erased \VNs can
be recovered using \gls{bp} decoding. In practice, to obtain
the threshold with a certain precision $\precision$, one fixes a
target erasure probability $\pe$ and a maximum number of iterations
$\lmax$. Then, starting from $\bar{\eps} = \precision$, one
iteratively computes \eqref{eq:de} until the average erasure
probability $(\sum_{j=1}^{L} p_j^{(l)})/L$ is either smaller than $\pe$
(successful decoding) or the number of iterations exceeds $\lmax$
(decoding failure). In the first case, $\bar{\eps}$ is increased by
$\precision$ until the decoding fails. For a given channel quality
parameter $\bar{\eps}$ up to the decoding threshold, we denote the
number of iterations until successful decoding by $\lsuccess$. 

We are interested in optimizing $\mathbf{A}$ in terms of the decoding
threshold for a given code ensemble. The baseline bit mapper realizes
a uniform random mapping of coded bits to channels. For this case,
we have $a_{i,j} = 1/m$ $\forall i,j$, and the corresponding
assignment matrix is denoted by $\Aunif$. To
establish the amount of threshold gain we can hope for by finding a
better $\mathbf{A}$, consider the following.  Each \gls{bec} has
capacity $C_i = 1-\eps_i$ and the average capacity is $\bar{C} =
\frac{1}{m} \sum_{i=1}^{m} C_i  = 1 - \bar{\eps}$,
where $1-\bar{\eps}$ would correspond to the capacity of a \gls{bec}
with erasure probability $\bar{\eps}$. By employing the baseline
mapper, however, the channel is effectively a \gls{bec} with erasure
probability $\bar{\eps}$ and the two-sided
$(\lDegree,\rDegree,L,\coupling)$ ensemble can approach the capacity
of this channel for appropriately chosen $(\lDegree,\rDegree)$ and $L
\to \infty$, $\coupling \to \infty$ \cite[Th. 10]{Kudekar2011}.
Hence, for very long chain length $L$ and smoothing parameter
$\coupling$, one would expect the potential gains in terms of
threshold improvement to be rather small.\footnote{This is in
agreement with the results reported for example in \cite{Cheng2012,
Nowak2012,Richter2007,Gong2011, Lei2009}, \IE when bit mappers are
optimized for good, capacity-approaching code ensembles the reported
gains over a uniform mapping are usually ``small''.} However, for
finite $L$ and $\coupling$, which is our main region of interest in
this paper, significant gains may still be possible. This is also an
important region for practical systems since increasing $L$ and
$\coupling$ leads to large block lengths (assuming a fixed and finite
$M$) and high decoding complexity. Moreover, for the same average
channel quality $\bar{\eps}$, an optimized bit mapper may significantly
reduce the number of decoding iterations until successful decoding
compared to the baseline bit mapper. Finally, these potential gains
come only at a small cost, \IE by replacing the baseline bit mapper. 


\section{Optimization}

Ideally, we would like to solve the problem 
\begin{align}
	\mathbf{A}_\text{opt} = \underset{\mathbf{A} \in \mathcal{A}^{m \times L}}{\text{argmax}} \quad &
	\bar{\eps}^*(\mathbf{A}). \label{eq:objective}
\end{align} 
It was already pointed out in \cite[Sec. IV]{Richardson2001a} that
directly optimizing a decoding threshold is difficult, simply due to
the fact that finding the threshold is computationally expensive. This
is especially pronounced for \gls{scldpc} codes and even for the
moderate chain lengths considered in this paper ($L \leq 40$), the
computational cost attached to one threshold computation is
significant in the context of an optimization routine. An alternative,
yet practical, approach, is to start with a certain channel quality
parameter $\bar{\eps}$ and then optimize the convergence behavior of
the ensemble in terms of decoding iterations. Then, one can calculate
the new threshold for the obtained assignment matrix and repeat the
whole procedure. Such an iterative approach was proposed in \cite[Sec.
IV]{Richardson2001a} to find optimized degree distributions
for irregular \gls{ldpc} codes. Based on this idea, we use the
following iterative optimization routine in order to find bit mappers
with good decoding thresholds. 


\begin{enumerate}
	\item Initialize the channel quality $\bar{\eps}$ to the decoding
		threshold for the baseline bit mapper, \IE $\bar{\eps} =
		\bar{\eps}^*(\Aunif)$.

	\item Find $\Aopt$ such that it minimizes the number of decoding
		iterations until convergence for the given channel quality
		$\bar{\eps}$, \IE
		\begin{align}
			\Aopt = \underset{\mathbf{A} \in \mathcal{A}^{m \times L}}{\text{argmin}} \quad
			\lsuccess. \label{eq:objective2}
		\end{align}
		To solve the optimization problem
		\eqref{eq:objective2}, we use differential evolution
		\cite{Storn1997}.

	\item For the found optimized $\Aopt$, calculate the new
		threshold $\bar{\eps}^*(\Aopt)$. If the threshold did not
		improve, stop. Otherwise, set $\bar{\eps} =
		\bar{\eps}^*(\Aopt)$ and go to step 2). 

\end{enumerate}

With this procedure, the computational complexity can be significantly
reduced. However, it is not guaranteed to be equivalent to a true
threshold optimization, \IE $\mathbf{A}_{\text{opt}} \neq \Aopt$ in
general.

\section{Results and Discussion}

In the following, we present optimization results assuming two
parallel \BECs according to Fig.~\ref{fig:channel_model}(a). We focus
on the two-sided and circular versions of the $(4,8,L,\coupling)$
ensemble, where $L \in \{10, 15, \ldots, 40\}$ and $\coupling \in \{2,
4\}$. Threshold values are computed assuming $\precision = 10^{-4}$,
$\pe = 10^{-6}$, and $\lmax = 5000$. For a code ensemble with design
rate $R$ and an assignment matrix $\mathbf{A}$, the gap to capacity is
defined as $\Delta \define 1 - \bar{\eps}^*(\mathbf{A}) - R$.  

\begin{figure}[t]
	\centering
	\subfloat[Threshold and Design Rate]{\includegraphics{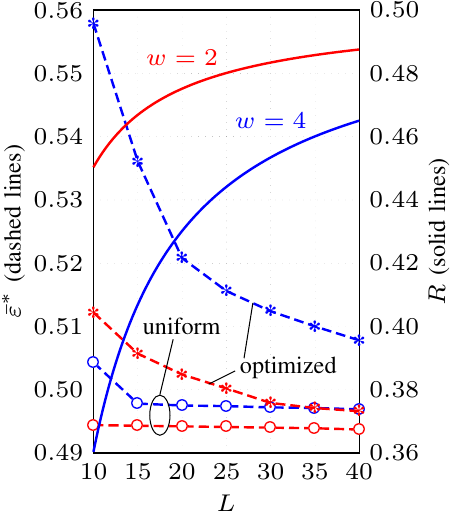}}
	\subfloat[Gap to Capacity]{ \includegraphics{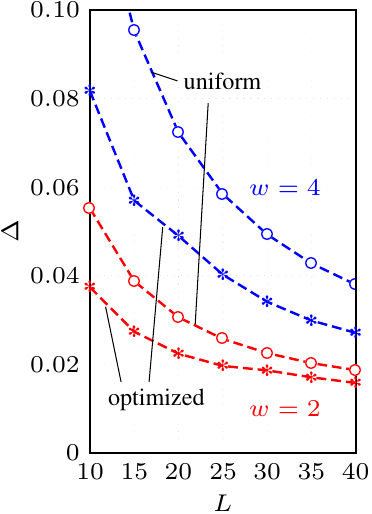}}

	\caption{Optimization results for the two-sided $(4,8,L,\coupling)$
	ensemble. Color is helpful.}
	\label{fig:gap}
\end{figure}

\subsection{Two-Sided Ensembles}

In Fig.~\ref{fig:gap}, the results for the two-sided ensembles are
shown. Fig.~\ref{fig:gap}(a) shows both the decoding
threshold (solid lines) and the design rate (dashed lines) and
Fig.~\ref{fig:gap}(b) shows the corresponding gap to capacity. The red and
blue lines correspond to $\coupling=2$ and $\coupling=4$,
respectively. 

Let us first briefly discuss the performance behavior for the
uniform baseline bit mapping schemes (circle markers),
which is essentially the same as for the case with only one \gls{bec}.
Both ensembles converge rapidly to a fixed threshold value for
increased chain lengths, \IE for $L\geq 15$ the decoding threshold is
approximately $0.497$ and $0.494$ for $\coupling = 2$ and $\coupling =
4$, respectively.\footnote{Due to the fixed maximum number
of iterations, \IE $\lmax = 5000$, the decoding threshold in fact
decreases slightly when $L$ increases.} Furthermore, the
ensemble with stronger coupling $(\coupling = 4)$ exhibits a
significantly larger rate loss. 

The thresholds that can be achieved with the optimized bit mappers are
shown by the star markers. It can be observed that it is possible to
improve over the ``uniform'' thresholds and, as expected, the absolute
threshold improvement becomes smaller with increasing chain length.
For $L \to \infty$, it was already mentioned that the expected
threshold gains will tend to zero. The results shown in
Fig.~\ref{fig:gap}(b) incorporate both the design rate and the
decoding threshold and allow for a comparison between the two coupling
parameters. We can conclude that a small coupling parameter is
beneficial in terms of $\Delta$, due to the large rate loss for
$\coupling = 4$ for both the uniform and optimized bit mappers.
Moreover, for a fixed gap to capacity, the optimized bit mappers allow
for a significant chain length reduction. As an example, for $\Delta =
0.04$ and $\coupling=4$, the chain length can be reduced from
approximately $L=40$ to $L=25$ and for $\Delta = 0.02$ and
$\coupling=2$ a chain length reduction from $L=35$ to $L=25$ is
possible. 

\begin{figure}[t]
	\centering
	\subfloat[$L=10$]{\includegraphics{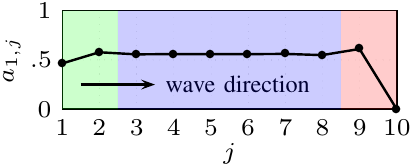}}
	\quad
	\subfloat[$L=20$]{\includegraphics{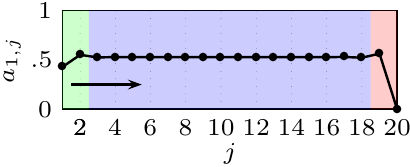}}

	\subfloat[$L=30$]{\includegraphics{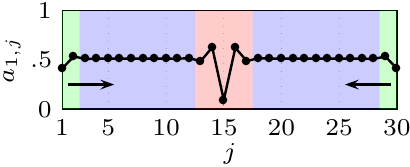}} \quad
	\subfloat[$L=40$]{\includegraphics{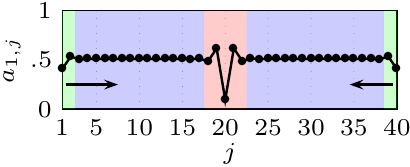}}
	
	\caption{Values in the first row of the optimized assignment matrices $\Aopt$ for
	the two-sided $(4,8,L,2)$ ensemble
	with different chain lengths.} 
	\label{fig:Aopt}
\end{figure}

\begin{figure}[t]
	\centering \subfloat[two-sided, $\bar{\eps}^*(\Aopt) = 0.5024$, $R =
	0.4752$, $\Delta = 0.0224$]{\includegraphics{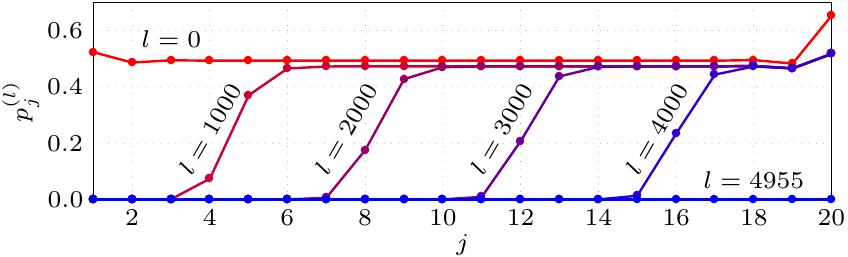}}

	\subfloat[circular, $\bar{\eps}^*(\Aopt) = 0.4772$, $R = 0.5$,
	$\Delta = 0.0228$]{\includegraphics{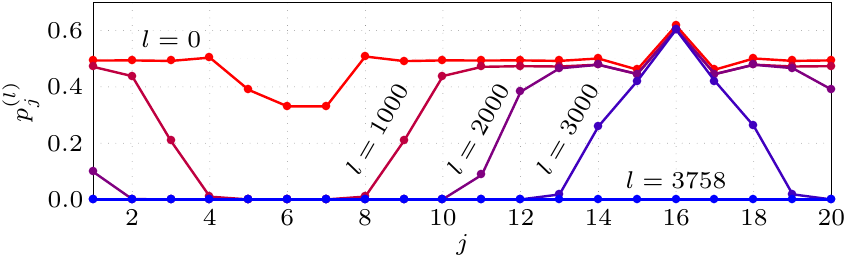}}

	\caption{Visualization of the iterative decoding behavior for (a)
	the two-sided and (b) the circular version of the $(4,8,20,2)$
	ensemble with optimized bit mappers. The optimized bit
	mapper induces a one-sided wave propagation for the two-sided
	ensemble in (a). }
	\label{fig:BP}
\end{figure}

Next, we show some of the found optimized bit mappers and discuss their
structure and the resulting iterative decoding behavior. In
Fig.~\ref{fig:Aopt}, the values in the first row of the optimized
assignment matrices $\Aopt$ are plotted for $\coupling = 2$ and
spatial lengths $L \in \{10, 20, 30, 40\}$. The values in
the first row (\IE $i = 1$) determine the fraction of \VNs at a
particular position to be sent over the \emph{good} channel, cf.\
Fig.~\ref{fig:channel_model}(a). Certain regions of the spatial
dimension are shaded in different colors and, when $\bar{\eps} =
\bar{\eps}^*(\Aopt)$, these regions correspond to the part where a
so-called decoding wave will start (green), end (red), and propagate
at a roughly constant speed (blue). First, let us focus on
Fig.~\ref{fig:Aopt}(a) to gain some insight into the general
structure of the optimized bit mappers. It is visible that the \VNs at
the last position are allocated only to the bad channel and there are
proportionally more \VNs allocated to the bad channel at the first
position (\IE $a_{1,1}$ is slightly less than $0.5$). For the
positions shaded in blue, the values of the assignment matrix are
roughly constant. Similar observations can be made for $L=20$. To
illustrate the effect of the optimized bit mappers, in
Fig.~\ref{fig:BP}(a) we provide a visualization of the iterative
decoding behavior for the two-sided $(4,8,20,2)$ ensemble at the
threshold value of $\bar{\eps}^*(\Aopt) = 0.5024$. As seen from the
figure, the optimized bit mapper induces a one-sided wave propagation
even though the ensemble is two-sided, and the wave propagates at a
roughly constant speed. The direction of the wave is
arbitrary due to the symmetry of the Tanner graph describing the
ensemble, \IE flipping each row in the assignment matrix leads to a
wave propagating from right to left with otherwise unchanged behavior.

The general structure of the optimized bit mappers for $L \in \{15,
25\}$ is similar to the ones shown in Fig.~\ref{fig:Aopt}(a) and (b).
For $L>25$, the structure changes as indicated in
Fig.~\ref{fig:Aopt}(c) and (d) for $L=30$ and $L=40$, respectively.
Here, the optimized allocation is such that two decoding waves
propagate from the ends of the spatial chain towards the center,
similarly as for a uniform bit mapper. The different structure
occurring for larger values of $L$ is possibly due to the fact that a
two-sided wave propagation leads to a faster convergence compared to
one-sided propagation for a given chain length $L$, even though a
one-sided propagation may be better in terms of threshold. However,
due to the fixed maximum number of decoding iterations, the iterative
optimization routine converges to these solutions for larger $L$. 

From the structure of the optimized bit mappers, an intuitive
explanation for the decoding threshold improvement can be given as
follows. In some sense, certain \gls{vn} classes are ``overprotected''
and proportionally more of the \VNs from these classes can be
allocated to the bad channel without harming the overall iterative
decoding performance. In turn, this allows for the remaining \gls{vn}
classes to be allocated more to the good channel, \IE channel uses
corresponding to the good channel become available and are spread out
evenly among the \VNs in the regions indicated in blue. In fact, all
optimized bit mappers for $\coupling = 2$ are such that in the blue
regions, the values for $a_{1,j}$ are all slightly greater than $0.5$.
From this one can also explain why asymptotically the threshold gain
will tend to zero, since asymptotically as $L \to \infty$ this effect
is not noticeable any more and $a_{1,j} \to 0.5$ in the blue regions. 

The optimized bit mappers for $\coupling = 4$ in principle show a
similar structure, but tend to be more unstable and wiggle around a
certain average value in the part of the spatial chain that is shaded
in blue. 


\subsection{Circular Ensembles} 

\begin{figure}[t]
	\centering
	\begin{minipage}{4.0cm}
		\subfloat[Gap to Capacity]{ \includegraphics{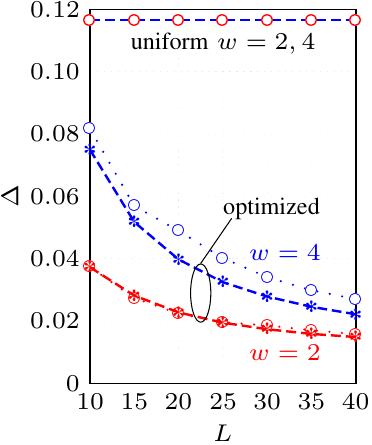}}
	\end{minipage}
	\quad
	\begin{minipage}{4cm}
	\subfloat[$L=20$]{\includegraphics{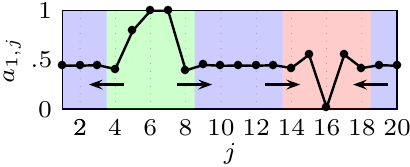}} 

	\subfloat[$L=40$]{\includegraphics{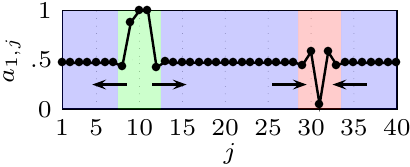}}
\end{minipage}

	\caption{Optimization results for the circular $(4,8,L,\coupling)$
	ensemble. In (b) and (c), we plot the values in the first row of the
	optimized assignment matrices for $\coupling = 2$. Color is
	helpful.}
	\label{fig:results_circ}
\end{figure}

In Fig.~\ref{fig:results_circ}, we show the optimization results for
the circular ensembles. In Fig.~\ref{fig:results_circ}(a), the
performance of the uniform and optimized bit mappers is shown in terms
of the gap to capacity. The dotted lines correspond to the performance of
the optimized bit mappers for the two-sided case and are simply
reproduced from Fig.~\ref{fig:gap}(b) for convenience to allow for a
comparison between the two-sided and circular ensembles. 
For circular ensembles, the design rate is $R = 1/2$, independent of
$L$, and the decoding threshold assuming a uniform bit mapper is given
by $\bar{\eps}^*(\Aunif) = 0.3834$ for both coupling parameters, hence
the constant ``uniform'' gap to capacity in
Fig.~\ref{fig:results_circ}(a). In fact, this threshold value
corresponds to the \gls{bp} decoding threshold of the regular,
uncoupled $(4,8)$ ensemble due to the absence of a termination
boundary. By employing the optimized bit mappers, a significant
threshold gain is possible, which directly translates into a
significant reduction in terms of the gap to capacity. Contrary to the
two-sided case, the threshold gain with respect to a uniform bit
mapper increases for longer chain lengths $L$.  Compared to the
results for the two-sided ensembles, one can achieve a slightly better
performance for $\coupling = 4$, while for $\coupling = 2$, the
corresponding curves in Fig.~\ref{fig:results_circ}(a) virtually
overlap. The improvement for $\coupling=4$ can be explained
by the large rate loss for the two-sided ensemble, and it shows that a
more careful design of the termination boundary for small $L$ may be
beneficial to achieve a better trade-off between rate loss and decoding
threshold performance.

Similarly as before, in Fig.~\ref{fig:results_circ}(b) and (c), we
show some of the optimized bit mappers in the form of the values in
the first row of $\Aopt$ for $\coupling =2$ and $L \in \{20, 40\}$.
The actual results from the
optimization routine have been adjusted (\IE appropriately flipped
and shifted), so as to make the figures look similar.  It can be
observed that for circular ensembles, the optimized bit mappers are
such that the \VNs over a small spatial range in the green region are
exclusively allocated to the good channel. In the red regions, the
optimized allocation resembles that of the two-sided ensembles for $L
\in \{30, 40\}$, cf.  Fig.~\ref{fig:Aopt}(c) and (d). The resulting
iterative decoding behavior is illustrated in Fig.~\ref{fig:BP}(b)
for the circular version of the $(4,8,20,2)$ ensemble at the decoding
threshold value of $\bar{\eps}^*(\Aopt) = 0.4772$. In essence, the
different channel qualities are exploited and a virtual termination
boundary is created by the optimized bit mapping schemes. This allows
for local convergence of the \gls{bp} decoder at these positions
within the first few iterations and consequently two waves propagate
outwards and eventually end in the region shaded in red. 


\subsection{Extension to More than Two Channels}

An extension to scenarios where more than two channels are present,
\EG for three \BECs as shown in Fig.~\ref{fig:channel_model}(b), is
straightforward but comes at the price of increased optimization
complexity due to the dimensionality increase of the problem. However,
once a good bit mapper for two \BECs is found, one can easily try the
following. Let $\Aopt \in \mathcal{A}^{2 \times L}$ be an optimized
assignment matrix for a given code ensemble of length $L$ and
$\bar{\eps}^*(\Aopt)$ the corresponding threshold. The input erasure
probabilities for the \VNs $(\eps^1, \dots, \eps^L)$ are thus fixed.
One can then try to find a feasible $\mathbf{A} \in \mathcal{A}^{3
\times L}$ that satisfies $\vect{\eps} \mathbf{A} = (\eps^1, \dots,
\eps^L)$, where $\vect{\eps} = (\eps_1, \eps_2, \eps_3)$ is a vector
with the individual erasure probabilities for $\bar{\eps} =
\bar{\eps}^*(\Aopt)$ in Fig.~\ref{fig:channel_model}(b). This can be
accomplished using standard numerical optimization routines. 


\section{Conclusion and Future Work}

In this paper, we studied \gls{scldpc} code ensembles over parallel
channels. Motivated by \gls{bicm}, we used an example with two
parallel \BECs and optimized the bit mapper that determines the
allocation of coded bits to the channels. Compared to a uniform
random bit mapper, the decoding threshold can be improved or,
alternatively, the spatial chain length can be reduced.  For circular
ensembles, the different qualities of the channels can be exploited to
obtain a wave-like decoding behavior similar to terminated, \EG
two-sided, ensembles.  
Future work includes the study of protograph-based ensembles for
finite length code design and the application to different channel
types. Further, it would be of much practical value to find an
analytical characterization of the optimal mappers.


\end{document}